\documentclass[11pt,dvips]{article}

\usepackage{epsfig,times} 
\usepackage{here}
\usepackage{amssymb}
\usepackage{wrapfig}
\usepackage{picinpar}
\makeatletter
\renewcommand{\section}{\@startsection{section}{1}{0in}
        {0.4\baselineskip}{0.1\baselineskip}{\Large\bf}}
\renewcommand{\subsection}{\@startsection{subsection}{2}{0in}
        {0.25\baselineskip}{-\baselineskip}{\large\bf}}
\renewcommand{\subsubsection}{\@startsection{subsubsection}{3}{0in}
        {0.1\baselineskip}{-\baselineskip}{\normalsize\bf}}
\makeatother
\newcommand{\icrc}{$26^{\rm th}$ ICRC\ }
\begin{document}

%
\makeatletter\newcommand{\ps@icrc}{
\renewcommand{\@oddhead}{\slshape{HE.6.2.02}\hfil}}
\makeatother\thispagestyle{icrc}
\markright{HE 6.2.2}
\begin{center}
{\LARGE \bf Formation of Centauro and Strangelets in Nucleus--Nucleus 
            Collisions at the LHC and their Identification by the ALICE
            Experiment}
           \footnote{
                     To appear in the proceedings of the 26th
                     International Cosmic Ray Conference,
                     17-25 August 1999, Salt Lake City.
                     \protect\newline\hspace*{4.5mm}
                     Further information at:
             http://home.cern.ch/${\sim}$angelis/castor/Welcome.html
                     }
\end{center}

\begin{center}
{\bf A.L.S. Angelis$^{1}$, J. Bartke$^{2}$, M.Yu. Bogolyubsky$^{3}$,
     S.N. Filippov$^{4}$, E. G\l{}adysz-Dziadu\'s$^{2}$,\\
     Yu.V. Kharlov$^{3}$, A.B. Kurepin$^{4}$, A.I. Maevskaya$^{4}$,
     G. Mavromanolakis$^{1}$, A.D. Panagiotou$^{1}$,\\
     S.A. Sadovsky$^{3}$, P. Stefanski$^{2}$ and Z. W\l{}odarczyk$^{5}$\\
}
{\it $^{1}$Division of Nuclear and Particle Physics, University of Athens,Greece.\\
     $^{2}$Institute of Nuclear Physics, Cracow, Poland.\\
     $^{3}$Institute for High Energy Physics, Protvino, Russia.\\
     $^{4}$Institute for Nuclear Research, Moscow,Russia.\\
     $^{5}$Institute of Physics, Pedagogical University, Kielce, Poland.\\
}
\end{center}

\begin{center}
{\large \bf Abstract\\}
\end{center}
\vspace{-0.5ex}
We present a phenomenological model which describes the formation
of a Centauro fireball in nucleus-nucleus interactions in the upper
atmosphere and at the LHC, and its decay to non-strange baryons and
Strangelets.
We describe the CASTOR detector for the ALICE experiment at the LHC.
CASTOR will probe, in an event-by-event mode, the very forward,
baryon-rich phase space 5.6 $ \le \eta \le $ 7.2 in 5.5$\times$A TeV
central $Pb + Pb$ collisions.
We present results of simulations for the response of the CASTOR
calorimeter, and in particular to the traversal of Strangelets.
\vspace{1ex}

\enlargethispage{8pt}
\section{Introduction:}
\label{sec:intro}

The physics motivation to study the very forward phase space in 
nucleus--nucleus collisions stems from the potentially very rich 
field of new phenomena, to be produced in and by an environment with
very high baryochemical potential.
The study of this baryon-dense region, much denser than the highest
baryon density attained at the AGS or SPS, will provide important
information for the understanding of a Deconfined Quark Matter (DQM)
state at relatively low temperatures, with different properties from
those expected in the higher temperature baryon-free region around
mid-rapidity, thought to exist in the core of neutron stars.

The LHC with an energy equivalent to $10^{17}$ eV for a moving proton
impinging on one at rest, will be the first accelerator to effectively
probe the highest energy cosmic ray domain. Cosmic ray experiments have
detected numerous most unusual events which have still not been understood.
These events, observed in the projectile fragmentation rapidity region,
may be produced and studied at the LHC in controlled conditions.
Here we mention the ``Centauro'' events and the ``long-flying component''.
Centauros (Lates, Fugimito \& Hasegawa 1980) exhibit relatively small
multiplicity, complete absence (or strong suppression) of the electromagnetic
component and very high $\langle p_{\rm T} \rangle$.
In addition, some hadron-rich events are accompanied by a strongly
penetrating component observed in the form of halo, strongly penetrating
clusters (Hasegawa \& Tamada 1996, Baradzei et al. 1992) or long-living
cascades, whose transition curves exhibit a characteristic form with many
maxima (Arisawa et al. 1994, Buja et al. 1982).

\section{A model for the formation of Centauro and Strangelets}
\label{sec:model}

A model has been developed in which Centauros are considered to originate
from the hadronization of a DQM fireball of very high baryon density 
$\rm (\rho_b \gtrsim 2~fm^{-3})$ and baryochemical potential
$\rm (\mu_b >> m_n)$, produced in ultrarelativistic nucleus--nucleus
collisions in the upper atmosphere (Asprouli, Panagiotou \&
G\l{}adysz-Dziadu\'s 1994, Panagiotou et al. 1992, Panagiotou et al. 1989).
In this model the DQM fireball initially consists of u, d quarks and gluons.
The very high baryochemical potential prohibits the creation of $\rm u\bar{u}$
and $\rm d\bar{d}$ quark pairs because of Pauli blocking of u and d quarks
and the factor exp~$\rm (-\mu_q/T)$ for $\rm \bar{u}$ and $\rm \bar{d}$ anti-
\newpage
\hspace*{-6mm}
quarks, resulting in the fragmentation of gluons into s$\bar{\rm s}$
pairs predominantly. In the hadronization which follows

\begin{wraptable}[24]{R}[0pt]{0pt}
\begin{tabular}{|c|c|c|}
\hline
\multicolumn{3}{c}{{\bf Table 1.} Average characteristic quantities of Centauro}\\
\multicolumn{3}{c}{events and Strangelets produced in Cosmic Rays and}\\
\multicolumn{3}{c}{expected at the LHC.}\\
\hline
  Centauro             &  Cosmic Rays             &  LHC                  \\
\hline
  Interaction          &  ``$Fe + N$''            & $ Pb + Pb $           \\
 $ \sqrt{s} $          & $ \gtrsim $ 6.76 TeV     &  5.5 TeV              \\
  Fireball mass        & $ \gtrsim $ 180  GeV     & $ \sim $ 500 GeV      \\
 $ y_{proj} $          & $ \geq $ 11              &   8.67                \\
 $ \gamma $            & $ \geq 10^4 $            & $ \simeq $ 300        \\
 $ \eta_{cent} $       &      9.9                 & $ \simeq $ 5.6        \\
 $ \Delta\eta_{cent} $ &      1                   & $ \simeq $  0.8       \\
 $ <p_T> $             &     1.75 GeV             &    1.75 GeV (*)       \\
  Life-time            & $ 10^{-9} $ s            & $ 10^{-9} $ s (*)     \\
  Decay prob.          & 10 \% (x $\geq$ 10 km)   & 1 \% (x $\leq$ 1 m)   \\
  Strangeness          &   14                     &  60 - 80              \\
 $ f_s $ (S/A)         & $ \simeq $ 0.2           & 0.30 - 0.45           \\
    Z/A                & $ \simeq $ 0.4           & $ \simeq 0.3 $        \\
 Event rate            & $ \gtrsim $ 1 \%         & $ \simeq $ 1000/ALICE-year \\
\hline
 ``Strangelet''        &  Cosmic Rays             &  LHC                  \\
\hline
  Mass                 & $ \simeq $ 7 - 15 GeV    & 10 - 80 GeV           \\
    Z                  & $ \lesssim $ 0           & $ \lesssim $ 0        \\
 $ f_s $               & $ \simeq $ 1             & $ \simeq $ 1          \\
\hline
\multicolumn{3}{r}{\small{(*) assumed}}
\end{tabular}
\end{wraptable}

\hspace*{-6.5mm}
this leads to the
strong suppression of pions and hence of photons, but allows kaons to be
emitted, carrying away strange antiquarks, positive charge, entropy and
temperature. This process of strangeness distillation transforms the initial
quark matter fireball into a slightly strange quark matter state.
The finite excess of s quarks and their stabilizing effects, in addition to
the large baryon density and binding energy and the very small volume, may
prolong the lifetime of the Centauro fireball, enabling it to reach
mountain-top altitudes (Theodoratou \& Panagiotou 1999).
In the subsequent decay and hadronization of this state non-strange
baryons and strangelets will be formed. Simulations show that strangelets
could be identified as the strongly penetrating particles frequently seen
accompanying hadron-rich cosmic ray events
(G\l{}adysz-Dziadu\'s \& W\l{}odarczyk 1997,
G\l{}adysz-Dziadu\'s \& Panagiotou 1995).

In this manner, both the basic characteristics of the Centauro events
(small multiplicities and extreme imbalance of hadronic to photonic
content) and the strongly penetrating component are naturally explained.
In table 1 we compare characteristics of Centauro and 
strongly penetrating components (Strangelets), either experimentally
observed or calculated within the context of the above model, for cosmic
ray interactions and for nucleus--nucleus interactions at the LHC.

\section{Design of the CASTOR detector}
\label{sec:det}

\begin{figwindow}[1,r,%
{\vspace*{-2mm}\mbox{\epsfig{file=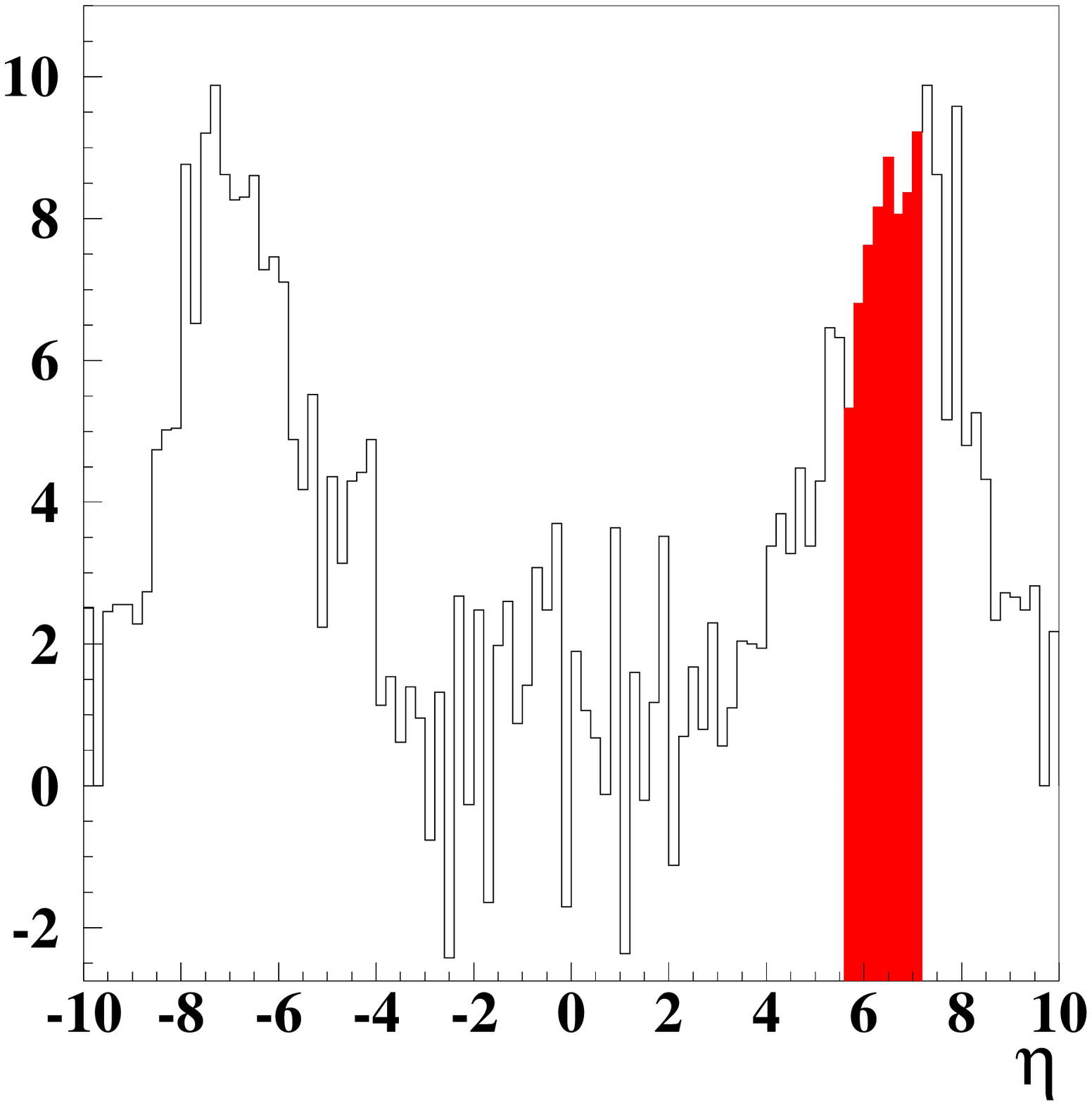,width=71mm}}},%
{Average net baryon number pseudorapidity distribution obtained from 50
central $Pb+Pb$ HIJING events.}{\label{fig:Nb_hij}}]
With the above considerations in mind we have designed the CASTOR
(Centauro And STrange Object Research) detector (Angelis et al. 1997)
for the ALICE heavy ion experiment at the LHC, in order to study the very
forward, baryon-dense phase space region.
CASTOR will cover the pseudorapidity interval $5.6 \le \eta \le 7.2$
and will probe the maximum of the baryon number density and energy flow.
It will identify any effects connected with these conditions and
will complement the physics program pursued by the rest of the ALICE
experiment in the baryon-free mid-rapidity region.
Figure~\ref{fig:Nb_hij} depicts the net baryon number pseudorapidity
distribution as predicted by the HIJING Monte-Carlo generator for an
average central $Pb+Pb$ collision at the LHC, with the acceptance of
CASTOR superimposed on the plot.

\hspace*{4mm}
The CASTOR calorimeter is azimuthally symmetric around the beam pipe
and is shown schematically in figure~\ref{fig:CastCal}.
It comprises electomagnetic and hadronic sections and is longitudi-
\end{figwindow}

\begin{figwindow}[1,r,%
{\mbox{\epsfig{file=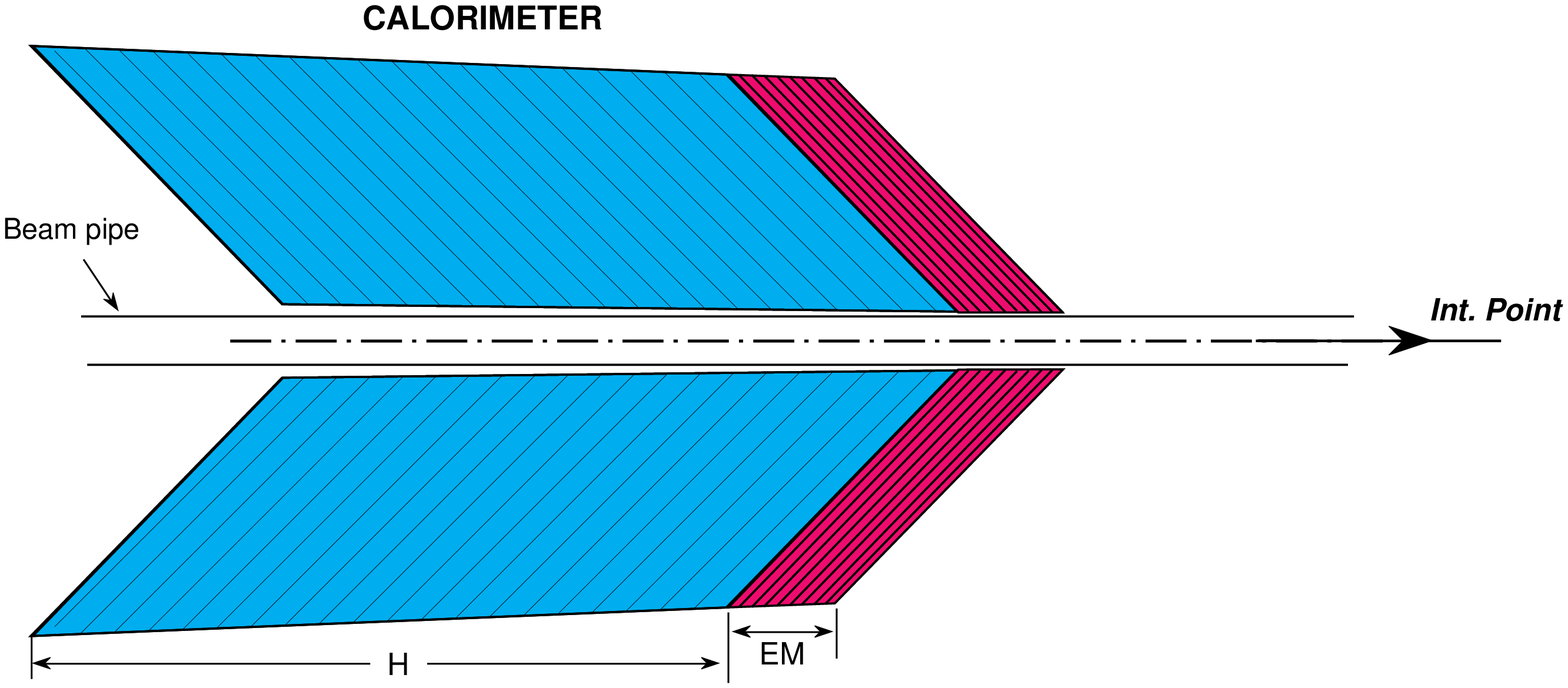,width=100mm}}},%
{Schematic representation of the CASTOR calorimeter.}{\label{fig:CastCal}}]
\hspace*{-6mm}
nally
segmented so as to measure the profile of the formation and propagation
of cascades. The calorimeter is made of layers of active medium sandwiched
between tungsten absorber plates. The active medium consists of planes of
silica fibres and the signal is the Cherenkov light produced as they are
traversed by the charged particles in the shower. The fibres are inclined at
45 degrees relative to the incoming particles in order to maximize the light
output. The calorimeter is azimuthally divided into 8 self-supporting octants.
In the current stage of the design each octant is longitudinally segmented
into 80 layers, the first 8 ($\simeq$ 14.7 X$_0$) comprising the
electromagnetic section and the remaining 72 ($\simeq$ 9.47 $\lambda_{\rm I}$)
the hadronic section. The calorimeter will be read out via air light guides
made out of stiff plastic, internally painted with UV reflecting paint.
The produced Cherenkov photons will propagate along the silica fibres to
the lateral surfaces of the calorimeter where they will exit into the light
guides. Inside the light guides they will be directed to PM tubes equipped
with quartz photocathode entry windows to optimally match the wavelength of
the Cherenkov light.
It is envisaged to couple together the light output from groups of 4
consecutive active layers into the same light guide, giving a total of 20
readout channels along each octant.
\end{figwindow}

\medskip
\section{Simulation of the CASTOR calorimeter performance}
\label{sec:simul}

\begin{figure}[H]
\begin{center}
\includegraphics[width=12cm]{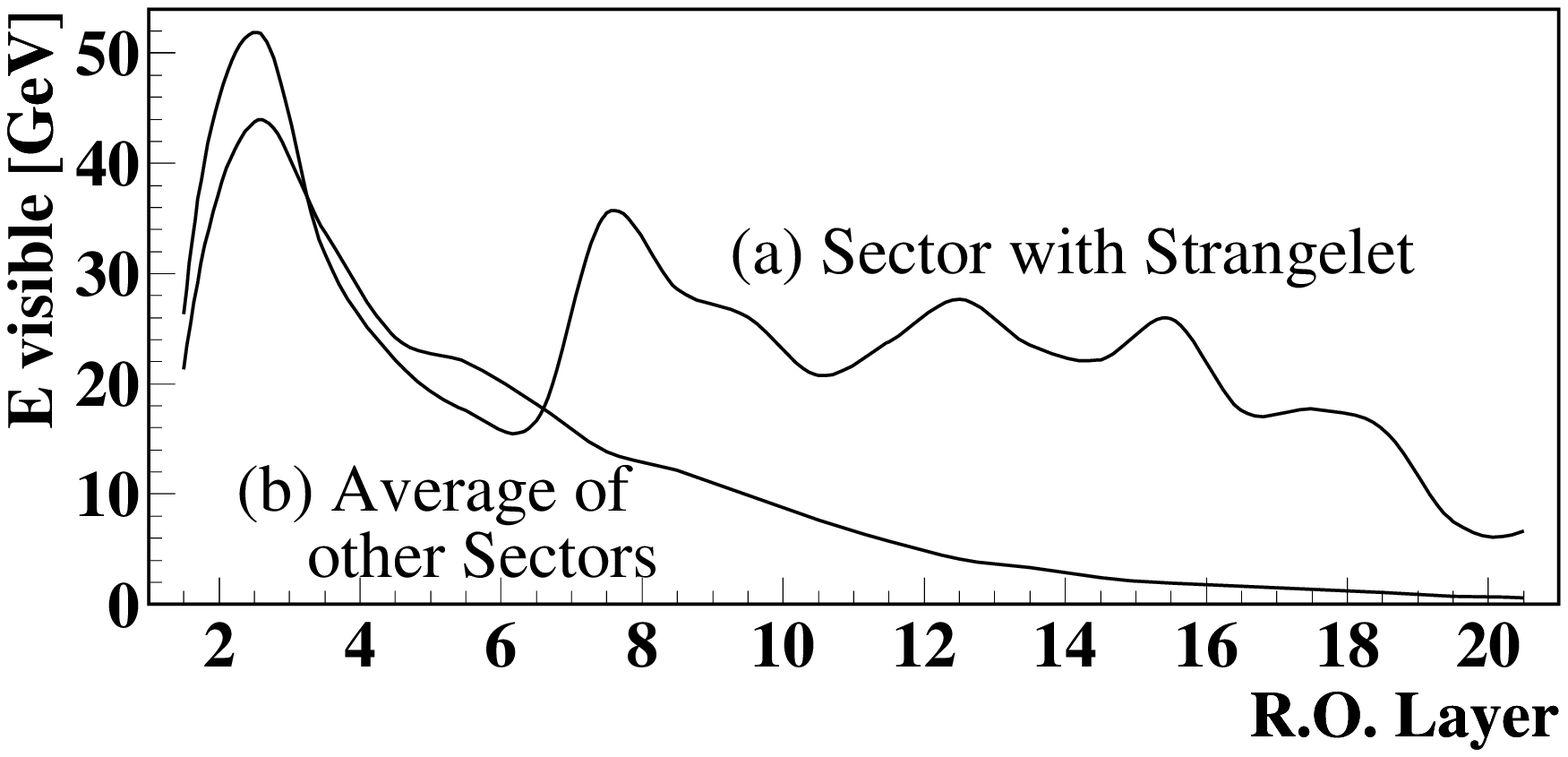}
\vspace*{-62mm}
\caption[]{Simulation of the energy deposition in the readout layers
           (couplings of 4 consecutive sampling layers) of the CASTOR
           calorimeter:
           (a) In the octant containing the Strangelet,
           (b) Average of the other octants.}
\label{fig:Slet}
\end{center}
\end{figure}

\begin{figwindow}[1,r,%
{\mbox{\epsfig{file=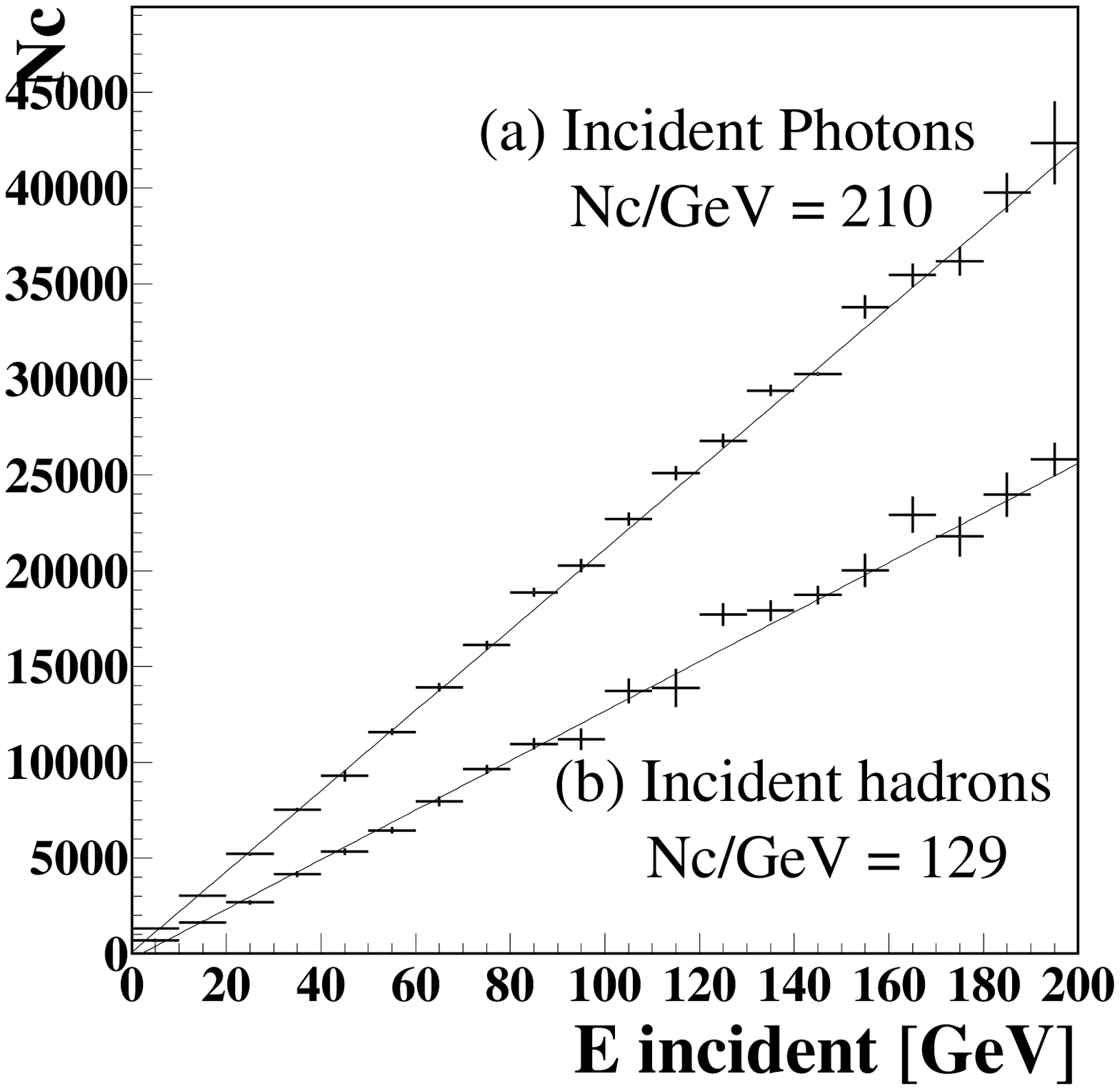,width=76.5mm}}},%
{Simulation of the total number of Cherenkov photons produced, retained
and propagated inside the fibres vs. incident particle energy:
(a) For incident photons, (b) For incident hadrons.}{\label{fig:Nc_vs_E}}]
We have made detailed GEANT simulations of the response of the CASTOR
calorimeter.
Figure~\ref{fig:Nc_vs_E} shows the total number of Cherenkov
photons produced, retained and propagated inside the fibres, as a function
of the incident particle energy for incident photons and hadrons from one
central LHC $Pb+Pb$ HIJING event. 
About 210 Cherenkov photons per GeV are obtained for incident photons and
129 Cherenkov photons per GeV for incident hadrons.
The accurate measurement of both the electromagnetic and hadronic energy
components is a prerequisite for an effective Centauro search and the
CASTOR calorimeter is optimized in that respect.

\hspace*{4mm}
In addition we have simulated the interaction of a Strangelet with the
calorimeter material, using the simplified picture described in (Angelis
et al. 1998, G\l{}adysz-Dziadu\'s \& W\l{}odarczyk 1997).
As an example figure~\ref{fig:Slet} shows the response of the calorimeter
to one central LHC $Pb+Pb$ HIJING event, to which has been added a Strangelet
of  $\rm A_{str}$=20, $\rm E_{str}$=20 TeV and quark chemical potential
$\rm \mu_{str}$=600 MeV (energy conservation has been taken into account).
Figure~\ref{fig:Slet}a shows the energy deposition along the octant
which contains the Strangelet,
while figure~\ref{fig:Slet}b shows the average of the energy deposition 
along the remaining seven octants.

\hspace*{4mm}
The study of such simulated events shows that the signal from an octant
containing a Strangelet is much larger than the average of the remainder,
while its transition curve displays long penetration and many maxima
structure such as observed in cosmic ray events.
\end{figwindow}

\newpage
\vspace*{-17mm}
\vspace*{-8mm}

\section{Conclusions}
\label{sec:concl}

We have developed a model which explains Centauro production in cosmic
rays and makes predictions for $Pb+Pb$ collisions at the LHC.
Our model naturally incorporates the possibility of Strangelet formation
and the ``long-flying component'' frequently seen accompanying hadron-rich
cosmic ray events is assimilated to such Strangelets.
We have designed a detector well suited to probe the very forward region
in $Pb+Pb$ collisions at the LHC, where very large baryon number density
and energy flow occur. CASTOR will identify any effects connected with
these conditions, while it has been particularly optimized to search for
signatures of Centauro and for long-penetrating objects.
We have simulated the passage of Strangelets through the CASTOR calorimeter
and we find large energy deposition, long penetration and many-maxima
structures similar to those observed in cosmic ray events.

\section*{Acknowledgements}
This work has been partly supported by the Hellenic General Secretariat
for Research and Technology $ \rm \Pi ENE \Delta $ 1361.1674/31-1/95,
the Polish State Committee for Scientific Research grants 2P03B 121 12
and SPUB P03/016/97,
and the Russian Foundation for Fundamental Research grant 96-02-18306.

\vspace{1ex}
\begin{center}
{\Large\bf References}
\end{center}
Angelis, A.L.S. et al. 1998, INP 1800/PH, Krak\'ow 1998.\\
Angelis, A.L.S. et al. 1997, CASTOR draft proposal, Note ALICE/CAS 97--07.\\
Arisawa, T. et al. 1994, Nucl. Phys. B424, 241.\\
Asprouli, M.N., Panagiotou, A.D. \& G\l{}adysz-Dziadu\'s, E. 1994,
Astropart. Phys. 2, 167.\\
Baradzei, L.T. et al. 1992, Nucl. Phys. B370, 365.\\
Buja, Z. et al. 1981, Proc. 17th ICRC, Paris Vol.11 p.104.\\
G\l{}adysz-Dziadu\'s, E. \& W\l{}odarczyk, Z. 1997,
J. Phys. G: Nucl. Part. Phys. 23, 2057.\\
G\l{}adysz-Dziadu\'s, E. \& Panagiotou, A.D. 1995,
Proc. Int. Symp. on Strangeness \& Quark Matter,
eds. G.~Vassiliadis et al., World Scientific, p.265.\\
Hasegawa, S. \& Tamada, M. 1996, Nucl. Phys. B474, 225.\\
Lates, C.M.G., Fugimito, Y.  \& Hasegawa, S. 1980, Phys. Rep. 65, 151.\\
Panagiotou, A.D. et al. 1989, Z. Phys. A333, 355.\\
Panagiotou, A.D. et al. 1992, Phys. Rev. D45, 3134.\\
Theodoratou, O.P. \& Panagiotou, A.D. 1999, \icrc .\\

\end{document}